\documentclass[%
 reprint,
 amsmath,amssymb,
 aps,
]{revtex4-2}

\usepackage{graphicx}% Include figure files
\usepackage{dcolumn}% Align table columns on decimal point
\usepackage{bm}% bold math
\usepackage{amsmath}
\usepackage{braket}
\usepackage{hyperref}
\usepackage{xurl}

\begin{document}

\preprint{APS/123-QED}

\title{Ferroelectric polarization controlled orbital Hall conductivity in a higher-order\\ topological insulator: \textit{d1T}-phase monolayer MoS$_2$}% Force line breaks with \\

\author{Yingjie Hu}
\author{Heng Gao$^{\ast}$}
\author{Wei Ren$^{\ast}$}

\affiliation{%
Physics Department, Materials Genome Institute, Shanghai Engineering Research Center for Integrated Circuits and Advanced Display Materials, Institute for Quantum Science and Technology, International Centre of Quantum and Molecular Structures, Shanghai University, Shanghai 200444, China
}%


\begin{abstract}
The higher-order topological insulator is an extended concept of the conventional topological insulator, which obeys the generalization of the standard bulk-boundary correspondence. In our paper, we predict the monolayer \textit{d1T}-phase transition metal dichalcogenide MoS$_2$ to be a higher-order topological insulator, while also possessing intriguing ferroelectric characteristics. We explicitly demonstrate the nontrivial topological index and reveal the hallmark corner states with quantized fractional charge within the bulk band gap. Second, we show the existence of a nonzero orbital Hall conductivity plateau within the energy gap which is a signature to identify higher-order topology system. Additionally, we investigate the relationship between the ferroelectricity and the orbital Hall conductivity of \textit{d1T} MoS$_2$ and find that the direction of ferroelectric polarization can modulate the positive and negative values of the orbital Hall conductivity $\sigma_{\rm{OH}}^x$. Our findings provide the theory and material candidate for ferroelectricity tunable orbital Hall effect which is promising to realize the external electric field controllable orbitronics.
\end{abstract}

\maketitle

%\tableofcontents

\section{Introduction}

Topological quantum materials have become one of the most important research fields in condensed-matter physics due to the novel and various quasiparticle with unique transport properties \cite{W1,W2,W3}. In the conventional $d$-dimensional topological insulators, from the bulk-boundary correspondence principle there exist $(d-1)$-dimensional conducting topological boundary states \cite{W4}. These boundary states are robust and protected time-reversal symmetry and characterized by Z$_2$ topological invariants \cite{W5,W6}. However, recent theoretical works have predicted a new class of higher-order topological insulator (HOTI) that obeys a generalization of the standard bulk-boundary correspondence \cite{W7}. HOTIs support topological states of two or more dimensions lower than the system. Specifically, a $d$-dimensional $n$th-order topological insulator is characterized by $(d-n)$-dimensional topological boundary states. For example, the second-order topological insulator possesses gapless corner states in two-dimensional systems or hinge states in three-dimensional systems. And the topological properties of HOTIs are protected by symmetries of the system, possibly augmented by the time reversal symmetry.

A series of HOTIs have been gradually discovered by first principles predictions \cite{W8,W9,W10,W11,W12,W13,W14,W15,W16,W60}, including the 2H phase MoS$_2$ \cite{W14}. Prior to the development of higher-order topological theory, this is a trivial topological insulator with a large band gap. However, it is found that triangular nanoflakes with armchair edges of 2H MoS$_2$ present in-gap corner states with fractional charge. So far, the material realization of second-order topological insulators in two-dimensional (2D) electronic systems is still rare. Therefore, proposing and discovering ideal and real material candidates of 2D HOTIs are urgent and important. In 2004, Murakami \textit{et al.} investigated finite-spin Hall conductivity (SHC) in conventional topological insulators PbTe and HgS \cite{W17}. Costa \textit{et al.} then found that within the energy gap of HOTIs, there exists a finite plateau of SHC, providing helpful insights for the discovery of new HOTIs \cite{W18}. Additionally, as the foundation of orbitronics, orbital Hall effect (OHE) recently gained renewed interest as the OHE has been revealed to serve as the fundamental origin of both spin and anomalous Hall Effect \cite{W19}. Recently a connection between orbital Hall conductivity (OHC) and HOTIs has also been discovered. Similarly to SHC, there exists a plateau of OHC the band gap of HOTIs \cite{W20}. 

The observation of OHE-induced orbital torques in magnetic materials leads to alternative approach for electrical control of magnetism \cite{W21,W22}. Hence, the realization of directional control of the OHE will bring breakthrough progress to its application in various fields.

In this work, we employ the density functional theory (DFT) to show that MoS$_2$ in the \textit{d1T}-phase is a new HOTI material. Its topologically protected gapless corner states have been found within a finite rhombic nanoflake. Additionally, we calculate both SHC and OHC of \textit{d1T} MoS$_2$, and find that only a significant OHC plateau exists within the band gap which can be used as a signature to identify this HOTI system. Moreover, we investigate the relationship between ferroelectric polarization and OHC, and successfully achieve directional control of the OHE by utilizing ferroelectric polarization flipping which offers a powerful method to manipulate the OHE in ferroelectric materials.

\begin{figure*}
  \includegraphics[width=0.95\textwidth]{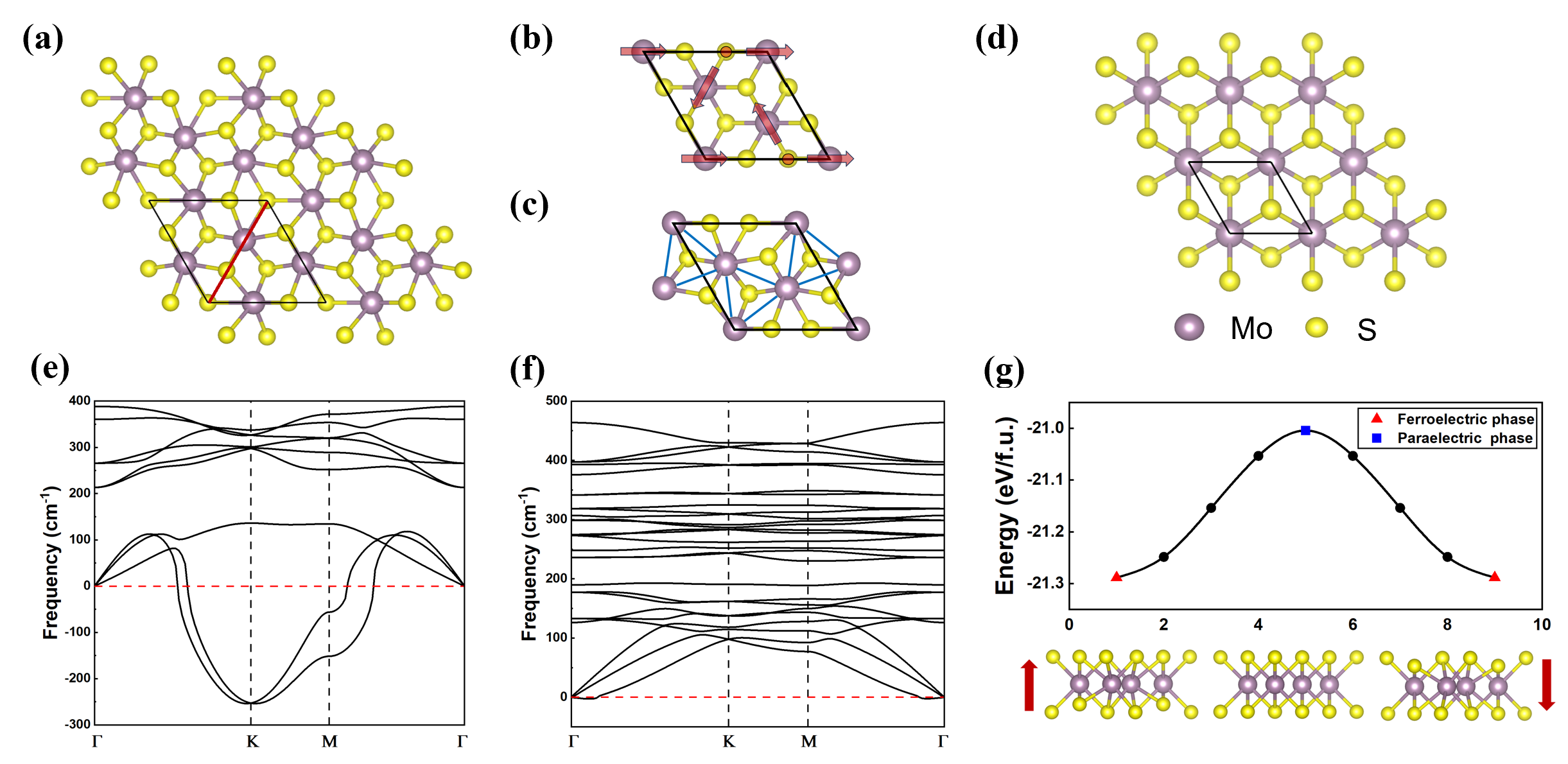}% Here is how to import EPS art
  \caption{\label{fig1}(a) Periodic structure of \textit{d1T} MoS$_2$ monolayer. The purple and yellow spheres represent Mo and S atoms, and the black lines represent its lattice. The red line represents the mirror symmetry axis. (b) Displacement vectors (depicted as red arrows) of the \textit{d1T} phase relative to the 1T phase. (c) The trimerization of Mo atoms in a $\sqrt{3}\times\sqrt{3}$ supercell. (d) Periodic structure of 1T MoS$_2$ monolayer. (e) The phonon spectra of 1T MoS$_2$ monolayer and (f) \textit{d1T} MoS$_2$ monolayer. (g) The transition state calculations for the reversible \textit{d1T} phase of MoS$_2$ with the 1T phase potential barrier of 0.284 (eV/f.u.).}
\end{figure*}

\section{Computational Methods}

The electronic structure calculations were performed by using DFT within the generalized gradient approximation of the Perdew-Burke-Ernzerhof (PBE) functional by the Vienna \textit{ab initio} simulation package (VASP) \cite{W23}. The cutoff energy is fixed to 500 eV, and all structures are relaxed until the energy and the forces are converged to 10$^{-7}$ eV and 0.0001 eV/Å, respectively. The Monkhorst-Pack $k$-point meshes used are 12$\times$12$\times$1 for sampling the Brillouin zone. A vacuum layer of 15 Å is used to avoid interactions between the nearest neighboring slabs. And the numerical calculations of SHC are performed by constructing the maximal localized Wannier functions (MLWFs) by Wannier90 package \cite{W24}. For the generation of MLWFs, we choose the $d$ orbitals of Mo and $p$ orbitals of S as the initial projections. The outer window and frozen energy window are used for the disentanglement procedure. In addition, the symmetry in the VASP calculation is switched off when calculating the data required for Wannier90. The OHC calculations are performed using our modified version of the open-source software WannierTools \cite{W25}. The tight-binding model is obtained using the Wannier90 code. And we have increased the sampling to 200$\times$200$\times$1 $k$ points in the 2D Brillouin zone to calculate the Hall conductivity.

\section{The crystal and band structures of \textit{d1T} MoS$_2$}

As is well known, monolayer MoS$_2$ has two widely studied structures: namely the hexagonal 2H phase and the octahedral 1T phase \cite{W28,W29}. The 2H phase is thermodynamically stable and has already been identified as a higher-order topological phase \cite{W14}. However, the centrosymmetric 1T phase exhibits instability at low temperature, revealed by DFT calculations with imaginary frequencies in the phonon spectrum, as shown in Fig.~\ref{fig1}(e). Several stable variant structures can be derived from the 1T structure, such as 1T$^{\prime}$, 1T$^{\prime\prime}$ and \textit{d1T} \cite{W30,W31} and our main focus here is on the \textit{d1T} structure. The Fermi nesting at $q\approx K$ point in 1T structure leads to a $\sqrt{3}\times\sqrt{3}$ supercell reconstruction \cite{W32}, which also results in the trimerization of Mo and breaks the central inversion symmetry, as depicted in Figs.~\ref{fig1}(b) and~\ref{fig1}(c). The space group of \textit{d1T} structure is $P31m$, featuring both $C_3$ and vertical mirror symmetries. The vertical mirror symmetry exists only along the shorter diagonal of the rhombic lattice, as shown in Fig.~\ref{fig1}(a). In addition, we calculate the phonon spectrum to check its stability and find no negative frequency as seen in Fig.~\ref{fig1}(f) , which means that the \textit{d1T} structure is dynamically stable.
\begin{figure}[h]
  \includegraphics[width=0.48\textwidth]{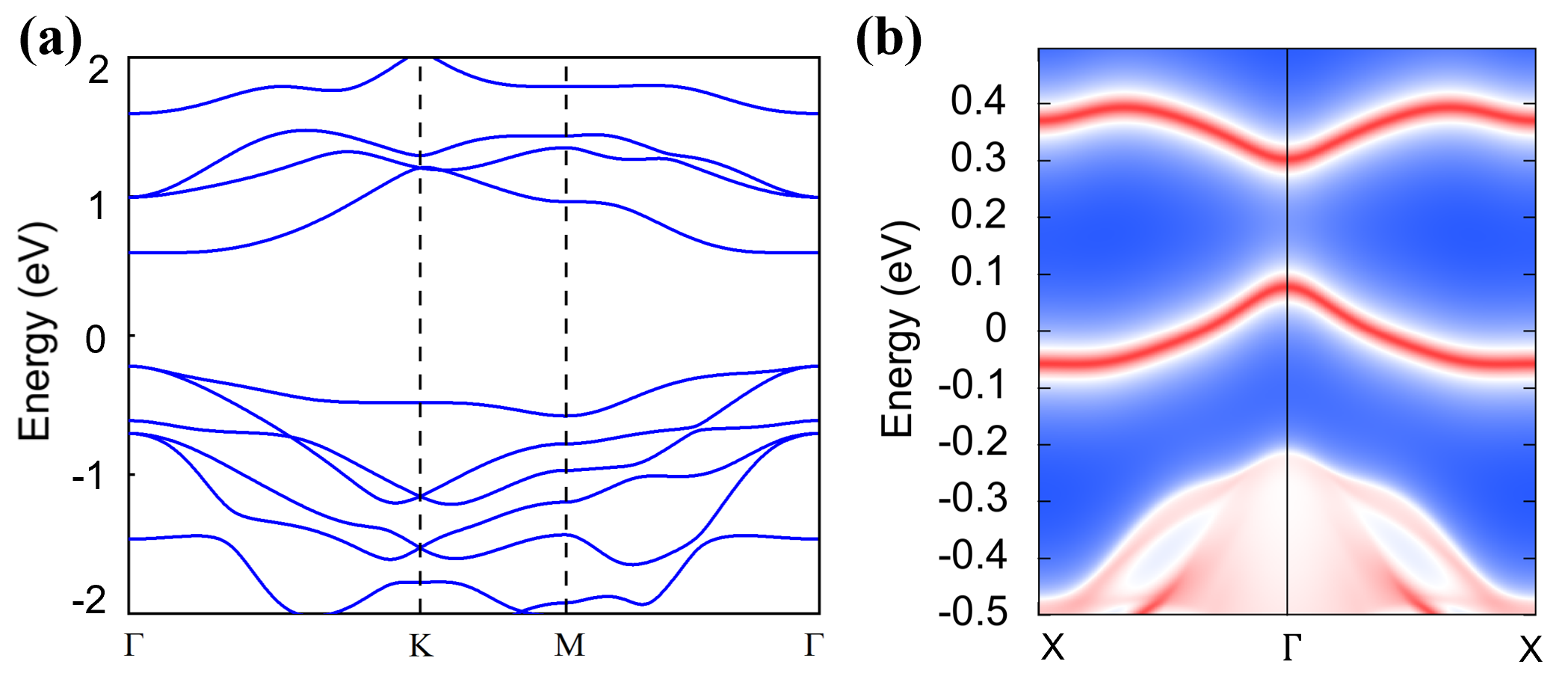}% Here is how to import EPS art
  \caption{\label{fig2} (a) The PBE band structure of \textit{d1T} MoS$_2$ monolayer. (b) The edge state of \textit{d1T} MoS$_2$ monolayer.}
\end{figure}

The \textit{d1T} structure is special in that it is a 2D ferroelectric material. Due to the breaking of the central inversion symmetry, an out-of-plane polarization is introduced. The direction of ferroelectric polarization is perpendicular to its 2D plane and points in the positive direction of the $z$ axis. And one can use the mirror symmetry operation in the $xy$ plane to construct a \textit{d1T} structure with the polarization direction reversed. In experiment, the out-of-plane ferroelectric polarization is switchable by the external electric field \cite{W33}. The barrier diagram of the ferroelectric polarization reversal is shown in Fig.~\ref{fig1}(g). We also calculated the band structure of this system without considering the spin-orbit coupling (SOC) effect, which indicates that it is an insulator with a PBE band gap of 0.85 eV, as shown in Fig.~\ref{fig2}(a).

To determine the topological properties, we calculate the edge states of \textit{d1T} MoS$_2$ monolayer, as illustrated in Fig.~\ref{fig2}(b). There exist two gaps, one gap is between the bulk and the edge states, and the other gap separates two edge states. This indicates that the system is not a conventional topological insulator, but has the possibility of a HOTI. Because as a HOTI, the topologically protected gapless states satisfy the bulk-boundary correspondence unless at least two dimensions lower than the bulk.

\section{Topological index and corner states of \textit{d1T} MoS$_2$ monolayer}

When a topological phase change occurs, i.e., the band gap closes and reopens, the symmetry eigenvalues change abruptly. Hence, we utilize symmetry indicators to characterize the bulk band topology. The bulk of transition metal dichalcogenides (TMD) with \textit{d1T} structure processes $C_3$ symmetry and there is no central inversion symmetry. According to Ref. \cite{W34}, the nontrivial topological classes arising from the $C_n$ symmetry can be distinguished through the symmetry representations that the occupied energy bands take at the high-symmetry point (HSP). For an HSP denoted by the symbol $\Pi$, the three-fold rotation symmetry eigenvalues should be $\Pi_n^{(3)}=e^{i2\pi(n-1)/3}$, with $n=1,2,3$. If the eigenvalues differ at different HSPs, then the energy band exhibits nontrivial topological property. We take the $\Gamma$ point as the reference point,

\begin{equation}
  \left[\Pi_n^{(3)}\right]=\#\Pi_n^{(3)}-\#\Gamma_n^{(3)},
\end{equation}

\noindent where $\#\Pi_n^{(3)}$ is the number of occupied bands with eigenvalue $\Pi_n^{(3)}$. For the trivial topological insulator, all the HSPs have the same symmetry eigenvalues. Thus, the trivial topological insulators have $\Pi_n^{(3)}=0$ for all the HSPs. For the $C_3$ symmetric HOTI, the topological index and fractional corner can be written in a form as

\begin{eqnarray}
  \chi^{(3)}=\left(\left[K_1^{(3)}\right],\left[K_2^{(3)}\right]\right),
  \\
  Q_{\rm{corner}}^{(3)}=\frac{e}{3}\left[K_2^{(3)}\right] \rm{mod}\ e,
  \label{eq:one}
\end{eqnarray}

\noindent where the superscript 3 of $\chi^{(3)}$ and $Q_{\rm{corner}}^{(3)}$ denotes the $C_3$ symmetry.

\begin{figure*}[htb]
  \includegraphics[width=0.9\textwidth]{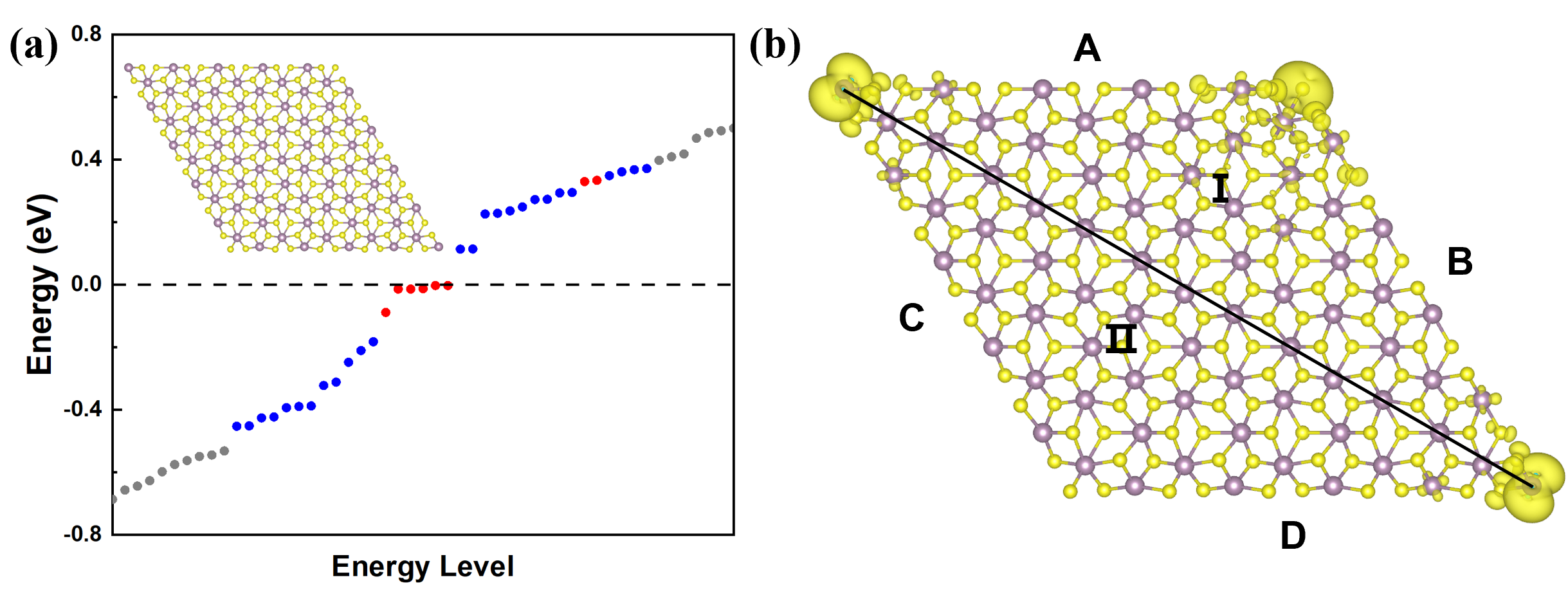}% Here is how to import EPS art
  \caption{\label{fig3} (a) The energy spectrum of the finite rhombic nanoflake of \textit{d1T} MoS$_2$. The gray, blue, and red dots stand for the bulk, edge, and corner states respectively. We focus on the corner states near the Fermi level with (b) the real-space distribution of six corner states in the middle of energy spectrum.}
\end{figure*}

The above topological index can be directly applied to the \textit{d1T} MoS$_2$. We use the irvsp package to calculate the symmetry eigenvalues of the occupied DFT energy bands \cite{W35}. For the monolayer, we obtain a nonzero topological index of $\left[-2,1\right]$, suggesting that this material is a potential HOTI with a fractional corner charge $Q_{\rm{corner}}^{(3)}=e/3$. For this 2D system, to prove its higher-order topological properties, it is necessary to find the hallmark corner states. We construct a finite rhombic nanoflake and calculate its energy spectrum based on the DFT calculation, as shown in Fig.~\ref{fig3}(a). This nanoflake is composed of 75 Mo and 150 S atoms which remains the original structure and the atomic ratio.

For the energy spectrum, it shows discontinuities near the Fermi level and there are several sets of degenerate states. We calculate the decomposed charge density of these states and plot their charge density distributions in real space. In the middle of the edge states, there exist six states whose real-space charge density distribution is plotted in Fig.~\ref{fig3}(a). Such six states are well located at three corners of the rhombic nanoflake and their positions in the energy spectrum are near the Fermi level which means they have appropriate electron filling. In this rhombic structure, $C_3$ symmetry is broken and mirror symmetry exists only on the diagonal of the two obtuse angles as seen the red solid line in Fig.~\ref{fig1}(a). Therefore, these corner states will only be symmetric at two acute angles, and the geometric constructions of A and B sides are different from the C and D sides as shown in Fig.~\ref{fig3}(b). Moreover, regions I and II divided by the long diagonal are also nonsymmetric with respect to it. Hence, it is normal for the corner states at the two obtuse angles to be distinctive. And we further identify the reason for the disappearance of this corner state: the charge originally localized at the corner diffuses to both sides, resulting in its transformation into an edge state, as shown in supplementary material Fig. S3 \cite{SUP}.

Through the above calculation of topological index and the discovery of corner states, we have determined that \textit{d1T} MoS$_2$ is a HOTI. Additionally, we attempt to correlate the higher-order topological properties of the \textit{d1T} system with consideration of the SHC or OHC.

\section{Spin and orbital Hall conductivity of HOTI \textit{d1T} MoS$_2$ monolayer}

The spin Hall effect (SHE) refers to a response of the spintronic system in the direction perpendicular to the applied electric field, which manifests as the spin current with no charge current but electrons of opposite spin moving in opposite directions \cite{W36}. The SHE can be separated into intrinsic and extrinsic parts \cite{W37}. In materials with strong SOC, the intrinsic spin Hall effect makes a substantial contribution to the total SHE, and it can also be accurately calculated using DFT calculation \cite{W26,W47,W48}. SHC has significant applications in the topological insulator system, where an integer multiple of the SHC plateau could exist in the band gap of a topological insulator \cite{W17}. Moreover, OHE refers to the transverse flow of orbital angular momentum (OAM) in response to a longitudinally external electric field \cite{W38,W39}. It resembles the SHE, except that the response of the system to the external electric field becomes an OAM current and it does not necessarily require SOC. According to the linear response theory, similar to the spin current, the intrinsic OAM current is time-reversal symmetric, and is governed by the Fermi sea term of the Kubo formula \cite{W20,W40}.

Recently, the connections between the SHC or OHC, and HOTIs have been found, where a finite SHC or OHC plateau emerges within the bulk band gap of HOTIs \cite{W18,W20}. This result expands the application of SHC and OHC in HOTIs. In this section, we calculate the SHC and OHC of \textit{d1T} MoS$_2$ through DFT calculation with an aim to implement the connection theory within our system, specifically employing SHC or OHC as a signature for identifying our HOTI system. The Kubo formula \cite{W41,W43,W46,W65} for OHC or SHC is given by

\begin{equation}
\sigma _{\rm{OH}(\rm{SH})}^{\eta}=\frac{e}{(2\pi)^2}\sum_{n}\int_{\rm{BZ}}d^2kf_{n,k}\Omega _{n,k}^{X_\eta}.
\end{equation}

\noindent where $\sigma _{\rm{OH}(\rm{SH})}^{\eta}$ is the orbital Hall (spin Hall) dc conductivity with ortbital (spin) polarization along the $\eta$ direction, and $\rm{BZ}$ is the Brillouin zone. 

\begin{equation}
  \label{1}
  \Omega _{n,k}=2\hbar \sum_{m\ne n}\mathrm{Im} \left[\frac{\bra{\psi_{n,k}}j_{y,k}^{X_\eta}\ket{\psi_{m,k}}\bra{\psi_{m,k}}v_x(k)\ket{\psi_{n,k}}}{(E_{n,k}-E_{m,k})^2}\right],
\end{equation}

\noindent represents the Berry curvature. Here $n$ and $m$ are band indices, and $E_{n,k}$ and $E_{m,k}$ denote the eigenvalues of the Hamiltonian in reciprocal space. The $f_{nk}$ is Fermi distribution function, and $v_i=\frac{1}{\hbar}\frac{\partial H(k)}{\partial k_i}(i=x,y)$ is the velocity operator. The current density operator component along $y$ with polarization $\eta$ is defined as $j_{y,k}^{X_{\eta}}=\left[X_{\eta}v_y(k)+v_y(k)X_{\eta} \right]/2$ \cite{W55,W68}. For the SHC, $X_{\eta}=s_{\eta}$ and for the OHC, $X_{\eta}=l_{\eta}$, $s_{\eta}$, and $l_{\eta}$ represent the $\eta$ components of the spin and orbital angular momentum operators, respectively.
\begin{figure*}[htb]
  \includegraphics[width=0.95\textwidth]{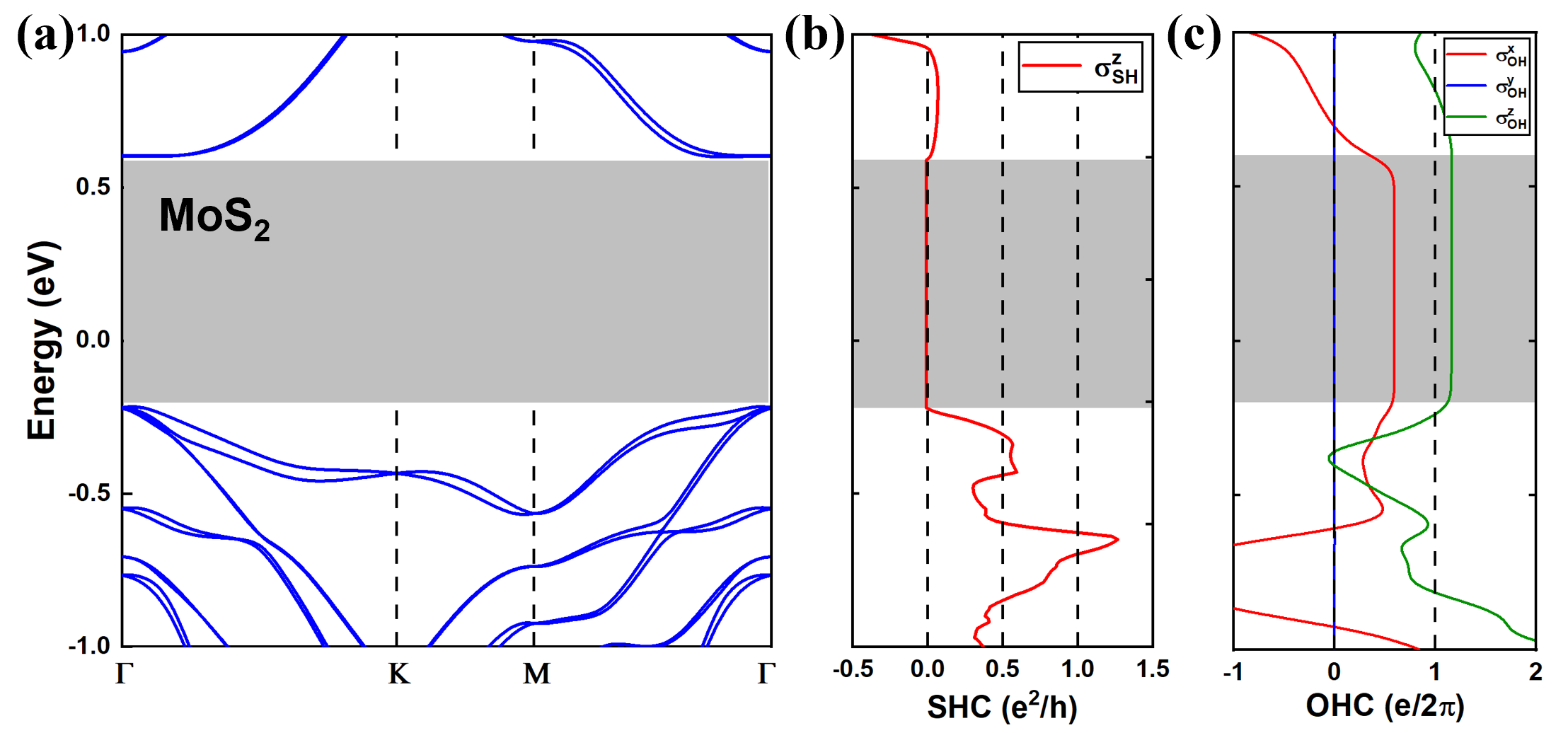}% Here is how to import EPS art
  \caption{\label{fig4} (a) The band structure of monolayer \textit{d1T} MoS$_2$. (b) The variation of SHC with respect to the position of Fermi energy for MoS$_2$. (c) The variation of OHC with respect to the position of Fermi energy for MoS$_2$.}
\end{figure*}

Figures~\ref{fig4}(a) and~\ref{fig4}(b) shows the band structure of \textit{d1T} MoS$_2$ and the variation of SHC with respect to the position of the Fermi energy. The results indicate that, for the \textit{d1T} MoS$_2$, there is no significant finite SHC within the band gap, akin to trivial topological insulators. This implies that the HOTI \textit{d1T} phase MoS$_2$ cannot be distinguished by the SHC signature, which is consistent with that the 2H-phase MoS$_2$ system being not identified as a HOTI by SHC either \cite{W20}. Hence, relying solely on the SHC as a hallmark may not suffice for identifying all HOTIs. Apart from that, it was reported to use the OHC to successfully identify the higher-order topology of 2H phase TMD. Therefore, we continue our identification by calculating the OHC of \textit{d1T} MoS$_2$, as shown in Fig.~\ref{fig4}(c).

We note that within the insulating gap of \textit{d1T} MoS$_2$, there exist OHC plateaus with values $\sigma_{\rm{OH}}^z=1.160$ and $\sigma_{\rm{OH}}^x=0.594$ in unit of $(e/2\pi)$. And the components $\sigma_{\rm{OH}}^y$ vanish. Following Ref. \cite{W49}, we define the absolute value of OHC as $|\sigma _{\rm{OH}}|=\sqrt{(\sigma _{\rm{OH}}^x)^2+(\sigma _{\rm{OH}}^y)^2+(\sigma _{\rm{OH}}^z)^2 }$. In our system, the absolute value of OHC is $1.303\ (e/2\pi)$. This is a relatively significant OHC platform, which indicates that the higher-order topological properties of \textit{d1T} structures can be identified by OHC. As we expected, the identification of the higher-order topology of the \textit{d1T} MoS$_2$ is consistent with the 2H phase which is that the higher-order topological phase can be witnessed by the OHE. 
%And we have proved that this conclusion can be directly applied in our newly discovered HOTI system.

\section{The manipulation of OHC by ferroelectric polarization switching}

\begin{figure*}[htb]
  \includegraphics[width=0.9\textwidth]{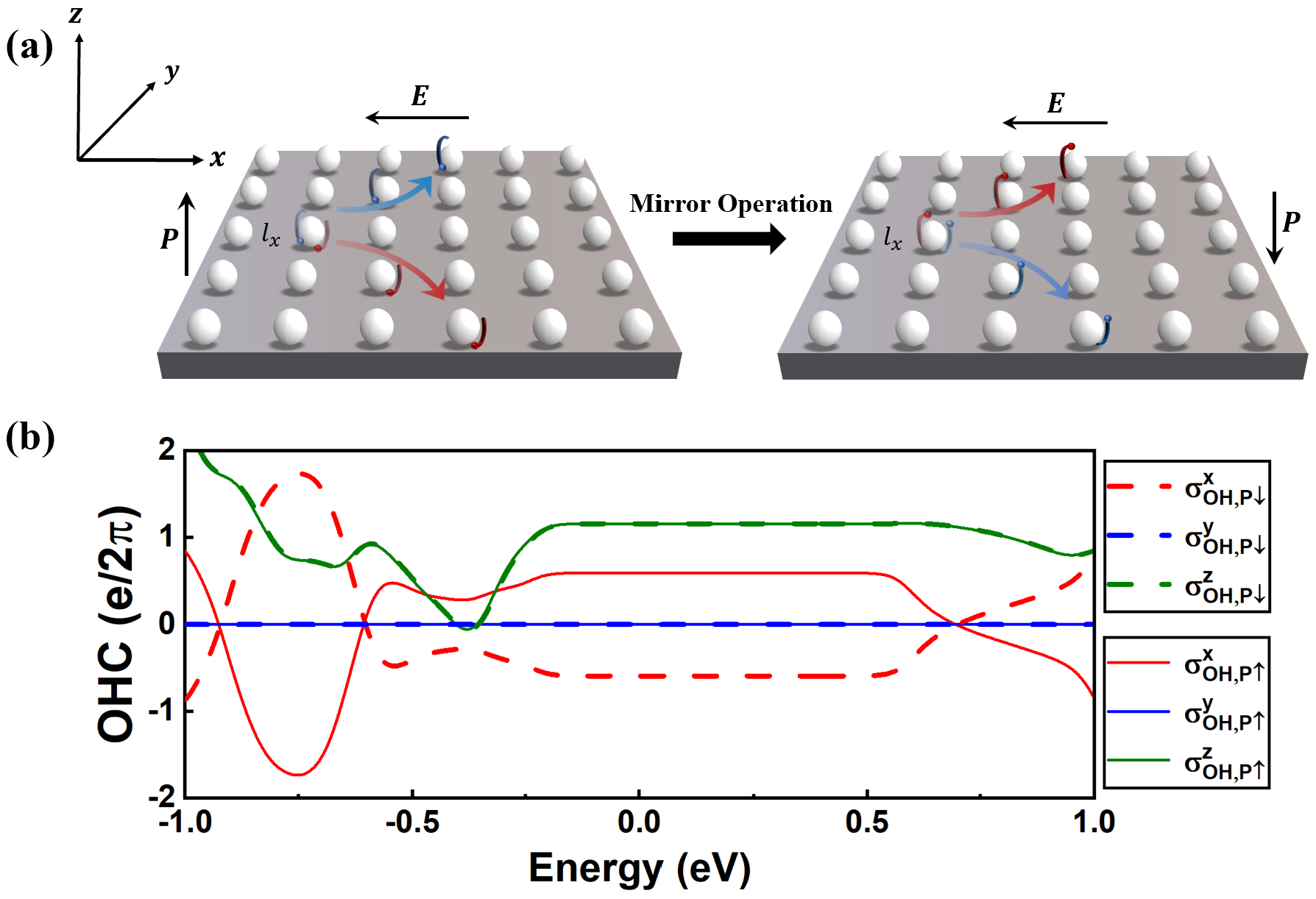}% Here is how to import EPS art
  \caption{\label{fig5} (a) Schematic diagram depicting the change in OAM $l_x$ under mirror symmetry in the $xy$ plane. (b) The variation of OHC with respect to the position of Fermi energy. The solid and dashed lines represent the OHC of the ferroelectric phase with polarization direction pointing upwards and downwards respectively.}
\end{figure*}

According to Ref. \cite{W50}, ferroelectricity and corner states are coupled together by crystallographic symmetry. Meanwhile, it is proved that in-plane polarization will lead to the emergence of corner states. However, the polarization direction of \textit{d1T} MoS$_2$ is along the positive $z$ axis which means that it is an out-of-plane ferroelectric polarization system. We then explore the relationship between the direction of ferroelectric polarization and OHC. A polarization-reversed \textit{d1T} structure is constructed through a mirror symmetry operation of the $xy$ plane, and we use the same method to calculate the OHC of the polarization-reversed structure, as shown in Fig.~\ref{fig5}. It can be found that $\sigma_{\rm{OH}}^y$ remains zero and $\sigma_{\rm{OH}}^z$ remains unchanged. Only the $x$ component $\sigma_{\rm{OH}}^x$ differs by one sign in the polarization up and polarization down structures which shows a possibility of using the out-of-plane direction of ferroelectric polarization to manipulate the OHC.

According to Ref. \cite{W45}, the SHC $\sigma_{yx}^y$ is not alowed in the symmetry group 157 ($P31m$). Therefore, similar to SHC, the $y$ component of OHC is also not allowed which remains zero in our calculations. And other results can be attributed to the alteration of Eq.~(\ref{1}) induced by the mirror symmetry operation. Under the mirror symmetry operation of the $xy$ plane, the $x$ and $y$ components of the velocity operator remain unchanged, while the $z$ component changes its sign. For the OAM operator in the orbital current, the components $l_x$ and $l_y$ change sign while $l_z$ remains unchanged. The variation of the velocity operator is straightforward to understand, and we will explain the change in the sign of OAM operator from a microscopic perspective. Let us consider the OAM operator as a vortex. With regards to $l_z$, the rotational direction of the vortex remains unchanged with the mirror symmetry of $xy$ plane. However, for $l_x$ or $l_y$, the rotational direction undergoes a shift, transitioning from clockwise to counterclockwise, or vice versa, resulting in a change of sign, as shown in Fig.~\ref{fig5}(a). Specifically, in our two systems with opposite polarization directions obtained by mirror symmetry, the OHC $\sigma_{xy}^z$ remains constant due to the unchanged sign of $v_x$, $v_y$, and $l_z$. And the $\sigma_{xy}^x$ of two systems differ by a sign because of $l_x$ changes.

In summary, we have achieved the modulation of OHC by employing ferroelectric polarization flipping in ferroelectric materials. Furthermore, we have elucidated the underlying mechanism of this modulation by analyzing the impact of mirror symmetry operations on our system.

\section{Conclusions}

In conclusion, we have discovered \textit{d1T} MoS$_2$ to be a new HOTI with the nontrivial higher-order topological phase recognized by a nonzero topological index, and its hallmark corner states are found in the finite rhombic nanoflake. Furthermore, our calculations of the SHC and OHC for \textit{d1T} MoS$_2$ reveal that only OHC exhibits a significant plateau within the band gap. Consequently, it becomes evident that our HOTI system can exclusively be distinguished through OHC measurement. Additionally, we explored the correlation between ferroelectric polarization and OHC. By manipulating the reversible directions of ferroelectric polarization, we have achieved the control of the positive and negative signs of specific components of OHC in the HOTI \textit{d1T} MoS$_2$.

\section{acknowledgment}

This work was supported by National Natural Science Foundation of China (Grants No. 12204299, No. 12074241, No. 12311530675, and No. 52130204), Science and Technology Commission of Shanghai Municipality (Grants No. 22XD1400900, No. 21JC1402700, No. 21JC1402600, and No. 22YF1413300), and High-Performance Computing Center, Shanghai Technical Service Center of Science and Engineering Computing, Shanghai University.

\nocite{*}
\email{gaoheng@shu.edu.cn}
\email{renwei@shu.edu.cn}
\bibliography{apssamp}% Produces the bibliography via BibTeX.

\end{document}